# Design of Matched Zero-Index Metamaterials Using Non-Magnetic Inclusions in Epsilon-Near-Zero (ENZ) Media

*Mário Silveirinha[1,2]and Nader Engheta[1]\**

*(1) University of Pennsylvania, Department of Electrical and Systems Engineering, Philadelphia, PA, U.S.A., engheta@ee.upenn.edu*

*(2) Universidade de Coimbra, Electrical Engineering Department – Instituto de Telecomunicações, Portugal,  mario.silveirinha@co.it.pt*

## Abstract

In this work, we study the electrodynamics of metamaterials that consist of resonant non-magnetic inclusions embedded in an epsilon-near-zero (ENZ) host medium. It is shown that the inclusions can be designed in such a way that both the effective permittivity and permeability of the composite structure are simultaneously zero. Two different metamaterial configurations are studied and analyzed in detail. For a particular class of problems, it is analytically proven that such matched zero-index metamaterials may help improving the transmission through a waveguide bend, and that the scattering parameters may be completely independent of the specific arrangement of the inclusions and of the granularity of the crystal. The proposed concepts are numerically demonstrated at microwaves with a metamaterial realization based on an artificial plasma.

---

\* To whom correspondence should be addressed: E-mail: engheta@ee.upenn.edu

# I. Introduction

The design and characterization of metamaterials is an important field of research today. These engineered complex materials may open new and exciting possibilities in various domains ranging from microwaves to optics, and may prospectively allow overcoming the diffraction limit in a number of problems (see e.g., [1]-[3]). Zero-index media are an interesting class of metamaterials formed by structures with index of refraction equal to zero (at the frequency of interest). At the infrared and optical frequencies, some low-loss semiconductors and dielectrics such as Silicon Carbide (SiC) may already have permittivity near zero, and are examples of zero-index materials available in nature [3]. Otherwise, they may be in principle synthesized as metamaterials. It has been suggested that zero-index materials can be used to narrow the far-field pattern of an antenna embedded in the medium, to transform curved wavefronts into planar ones, and to design delay lines [4]-[5]. Also recently, we proposed using ENZ materials to enhance the efficiency of some waveguiding devices and reduce the reflection coefficient at a junction or bend [6]. In general, unless both the permittivity and permeability are simultaneously zero, a zero-index material is not matched to free-space and so the reflectance may be very high. A noteworthy exception to this rule may occur if one physical dimension of the material is electrically small [6]. However, that restriction may be too severe for certain applications and so it is important to investigate the possibility of designing zero-index media matched with free-space. These materials have both $\varepsilon$ and $\mu$ equal to zero, and their electrodynamics was theoretically investigated in [4]. In this paper we will discuss how these structures can be designed as metamaterials at some frequency by embedding non-magnetic inclusions within an ENZ host medium, and study the transmission of an electromagnetic wave through a block of such metamaterials. Notice that previous works [7]-[8] have demonstrated the emergence of artificial magnetism in metamaterials with plasmonic inclusions embedded in a regular dielectric, and explored that effect to design left-handed materials. However, here we aim at matched zero-index metamaterials.

This paper is organized as follows. In section II we study the homogenization problem and derive closed-form formulas for the effective parameters of a metamaterial with an ENZ host medium. In section III we apply the results to two specific configurations: a periodic medium formed by circular rods and a periodic medium formed by non-uniform rings. In section IV, the response of a finite sized sample of the metamaterial is investigated in a waveguide scenario. It is demonstrated that when the permittivity and permeability of the material are matched, the granularity of the structure is not seen by an incoming wave. A metamaterial realization is proposed at microwaves based on the concept of artificial plasma. In section V the conclusions are drawn.

## II. The Homogenization Problem

To begin with, we consider that a non-magnetic material with permittivity near zero is readily available. As discussed before this may be the case at infrared and optical frequencies; ahead it will be explained how such media can be emulated at microwaves. We will use this material as our host medium (with permittivity $\varepsilon_h$) and we will design the non-magnetic inclusions in such a way that the effective permeability of the structure is zero. For simplicity we consider a 2D problem: the inclusions are uniform along the z-direction and the electromagnetic fields are H-polarized (magnetic field is $\mathbf{H} = H_z \hat{\mathbf{u}}_z$ with $\hat{\mathbf{u}}_z$ being the unit vector in the z direction, and $\partial/\partial z = 0$). The time variation is assumed to be $\exp(-i\omega t)$, where $\omega$ is the angular frequency. Thus, the electric field is given by $\mathbf{E} = (\nabla H_z / -i\omega\varepsilon) \times \hat{\mathbf{u}}_z$.

The geometry of the unit cell $\Omega$ of the periodic crystal is shown in Fig. 1. The host medium is assumed connected with permittivity near zero, $\varepsilon_h(\omega_p) \approx 0$, at the frequency of interest $\omega = \omega_p$, and the permittivity of the basic non-magnetic inclusion is $\varepsilon_i$ (not necessarily uniform). The boundary of the inclusion is denoted by $\partial D$, and the area of the unit cell is $A_{cell}$. Since the materials are assumed to be non-magnetic, we have $\mu = \mu_0$.

In [6] we proved that the electrodynamics of a two-dimensional medium with an ENZ component is very peculiar. In particular, we derived a fundamental property that establishes that in order that the electric field can be finite in the ENZ medium ($\varepsilon = \varepsilon_h$), it is necessary that $\nabla H_z = 0$ in the ENZ material, and consequently the magnetic field must be constant in the material:

$$H_z = H_z^{ext} = const. \text{ in the ENZ host medium} \tag{1}$$

We will use this fundamental property ahead. The objective next is to homogenize the metamaterial around the frequency $\omega = \omega_p$. First, we will study the permeability problem. By definition, the effective permeability (along z) of the periodic medium is given by,

$$\mu_{eff} = \mu_0 \frac{\langle H_z \rangle}{H_{bulk}} \tag{2}$$

where $\mu_0 \langle H_z \rangle$ is the average induction field (over the unit cell), $H_{\text{bulk}} = \langle H_z \rangle - m_z / A_{\text{cell}}$ is the "macroscopic" magnetic field, and $m_z$ is the magnetic dipole moment of the inclusion (per unit of length). We follow the classic homogenization approach of defining the effective parameters of a periodic medium from the properties of the corresponding electromagnetic Floquet-Bloch modes, i.e. solutions $(\mathbf{E}, \mathbf{H})$ such that $(\mathbf{E}, \mathbf{H}) e^{-i\mathbf{k}.\mathbf{r}}$ is periodic where $\mathbf{k} = (k_x, k_y, 0)$ is the wave vector.

A formal description of the nature of the Floquet modes at $\omega = \omega_p$ is presented in Appendix A. The main result is that besides the usual set of k-periodic modes, the electromagnetic crystal supports a set of *generalized* non-periodic Floquet modes with electric field of the form $\mathbf{E} = \dfrac{x_i}{a} \mathbf{E}_{p0} + \mathbf{E}_{p1}$, where $x_i$ is the coordinate along a generic direction of space, $\mathbf{E}_{p0}$ is a longitudinal periodic mode (see Appendix A), $\mathbf{E}_{p1}$ is a periodic function, and $a$ is a characteristic dimension of the crystal (e.g. the lattice constant). The emergence of such generalized modes, which to our best knowledge is overlooked in the literature, can be easily understood from a mathematical point of view as a degeneracy phenomenon at $\mathbf{k} = 0$. Indeed, consider a 1-dimensional wave $\psi(x) = A_{p0}(x; k) e^{ikx}$ propagating in a 1-dimensional reciprocal medium with $k = k(\omega)$ the Floquet constant, where $A_{p0}$ is the periodic amplitude of the wave. Let us also suppose that at some frequency $\omega = \omega_p$ the propagation constant vanishes, $k(\omega_p) = 0$, as well as the group velocity. Since the medium is reciprocal, as $\omega \to \omega_p$ the two modes that propagate along the positive and negative x-axis become more and more similar, and eventually at $\omega = \omega_p$ they collapse into the same mode. To lift this degeneracy a new non-periodic generalized mode proportional to $\dfrac{\partial \psi}{\partial k}$ emerges at $\omega = \omega_p$. Note that $\dfrac{\partial \psi}{\partial k}$ can be written as $\dfrac{\partial \psi}{\partial k}\bigg|_{k=0} = ixA_{p0}(x;0) + A_{p1}(x)$ where $A_{p1}$ is some periodic function. In a 2-D and 3-D problems a similar phenomenon occurs and as a consequence generalized Floquet modes may emerge at the point $\mathbf{k} = 0$.

The magnetic properties of a metamaterial are intrinsically related with the generalized Floquet modes. Before we present the details, it is important to note that even thought the electric field, $\mathbf{E} = \dfrac{x_i}{a}\mathbf{E}_{p0} + \mathbf{E}_{p1}$, associated with a generalized Floquet mode is non-periodic, the corresponding magnetic field is always periodic. This follows from the fact that $\mathbf{E}_{p0}$ is longitudinal, i.e. $\nabla \times \mathbf{E}_{p0} = 0$ (see Appendix A). In particular, the existence of generalized Floquet modes is completely consistent and compatible with the fundamental property mentioned earlier that implies that $H_z = H_z^{ext} = const.$ in the host medium at $\omega = \omega_p$.

In order to calculate the effective permeability we need to evaluate (2) for electromagnetic modes $(\mathbf{E}, \mathbf{H})$ around $\mathbf{k} = 0$. First we calculate the average induction field $\mu_0\langle H_z\rangle$. Applying Faraday's law to the boundary $\partial\Omega$ of the unit cell, it is readily found that:

$$V_\Omega \equiv \oint_{\partial\Omega} \mathbf{E}.\mathbf{dl} = i\omega\mu_0 \langle H_z\rangle A_{cell} \tag{3}$$

where $V_\Omega$ is the electromotive force across the boundary of the cell. In case $(\mathbf{E}, \mathbf{H})$ is associated with a periodic mode, it is clear from the definition that $V_\Omega = 0$ and consequently that the average induction field $\mu_0 \langle H_z \rangle$ also vanishes. Thus, periodic Floquet modes cannot be used to calculate the effective permeability using (2) (it can also be verified that $H_{\text{bulk}} = 0$ for periodic modes). Quite differently, for generalized modes $V_\Omega$ does not vanish because the electric field is not periodic. Thus, as hinted before, the magnetic properties of the metamaterial are defined by the generalized modes.

Next, we use the fact that $\varepsilon_h = 0$ at the frequency of interest, to find that the magnetic dipole moment (per unit of length) of an inclusion with cross-section $D$ (see Fig. 1) and centered at position $\mathbf{r}_i$ is given by:

$$
\begin{aligned}
\mathbf{m}_i &= \frac{-i\omega}{2} \int_D (\mathbf{r} - \mathbf{r}_i) \times (\varepsilon - \varepsilon_h) \mathbf{E} \, ds \\
&= \frac{1}{2} \int_D (\mathbf{r} - \mathbf{r}_i) \times (\nabla H_z \times \hat{\mathbf{u}}_z) \, ds \\
&= \hat{\mathbf{u}}_z \left( -\frac{1}{2} \int_{\partial D} H_z \, \hat{\mathbf{n}} . (\mathbf{r} - \mathbf{r}_i) \, dl + \int_D H_z \, ds \right)
\end{aligned}
\tag{4}
$$

In the above, the second identity is a consequence of Green's formula, $\partial D$ is the boundary of the inclusion, and $\hat{\mathbf{n}}$ is the unit normal vector oriented to the exterior. Remembering that $H_z = H_z^{ext} = const.$ in the host medium, and noting that this identity must also hold at the interface $\partial D$, we easily obtain that:

$$
m_z = -H_z^{ext} A_D + \int_D H_z \, ds
\tag{5}
$$

where $A_D$ is the area of the inclusion. In particular it is found that the magnetic dipole moment is independent of $\mathbf{r}_i$. This property is unusual and specific of the metamaterial under study. Indeed, it is well known that except in the static limit the magnetic dipole moment of an object depends on the origin of the coordinate system.

Next, using again (1) and the definition of $\langle H_z \rangle$ it is clear that (5) implies that:

$$m_z = \left(-H_z^{ext} + \langle H_z \rangle\right) A_{cell} \qquad (6)$$

Substituting this formula in the definition of $H_{\text{bulk}}$ we obtain the important result,

$$H_{\text{bulk}} = H_z^{ext} \qquad (7)$$

i.e. the (macroscopic) bulk magnetic field in the homogenized crystal is the same as the (microscopic) magnetic field in the ENZ (exterior) host region. Finally, substituting (3) and (7) into (2) we obtain that:

$$\mu_{eff} = \frac{V_\Omega}{i\omega A_{cell} H_z^{ext}} \qquad (8)$$

This shows that the effective permeability can be computed by calculating $V_\Omega$ and $H_z^{ext}$ for a generalized Floquet mode. In what follows we prove that this definition is self-consistent, i.e. it is independent of the considered mode. Moreover, we will explain how to formally obtain a closed analytical expression for $\mu_{eff}$.

To this end, we note that the magnetic field inside the inclusion is the solution of,

$$\nabla \bullet \left(\frac{1}{\varepsilon_i} \nabla H_z\right) + \omega^2 \mu_0 H_z = 0, \qquad (9)$$

subject to the Dirichlet boundary condition $H_z = H_z^{ext}$ at the boundary $\partial D$ (this is a consequence of (1)). The electric field in the interior of the inclusion is $\mathbf{E}^{\text{int}} = \left(\nabla H_z / j\omega \varepsilon_i\right) \times \hat{\mathbf{u}}_z$. Note that the electromagnetic fields inside the inclusion can be calculated without any knowledge of the electric field distribution outside. In particular, one can calculate the following characteristic parameter:

$$\overline{Z}_{int} = \frac{1}{A_{cell}} \frac{-\oint_{\partial D} \mathbf{E}^{int} . \mathbf{dl}}{H_z^{ext}} \tag{10}$$

We will refer to $\overline{Z}_{int}$ as the internal impedance of the inclusion (unities are $\Omega/m$). We stress that $\overline{Z}_{int}$ only depends on the geometry and electric properties of the inclusion. It is completely independent of the specific geometry of the crystal (apart from the normalization factor $1/A_{cell}$). In order to relate the effective permeability of the crystal with $\overline{Z}_{int}$, we apply Faraday's law to the domain $\Omega - D$ (host medium). Using (1) and (3) it is clear that,

$$V_{\Omega} - \oint_{\partial D} \mathbf{E}^{int} . \mathbf{dl} = i\omega\mu_0 A_{h,cell} H_z^{ext} \tag{11}$$

where $A_{h,cell}$ is the area of the host region. Substituting (10) and (11) into (8), we finally obtain that:

$$\mu_{eff} = \frac{\overline{Z}_{eq}}{-i\omega}, \qquad \text{with } \overline{Z}_{eq} \equiv \frac{-1}{A_{cell}} \frac{V_{\Omega}}{H_{z,ext}} = -i\omega\mu_0 \frac{A_{h,cell}}{A_{cell}} + \overline{Z}_{int} \tag{12}$$

The previous formula establishes that $\mu_{eff}$ is univocally determined by the internal impedance $\overline{Z}_{int}$. We will see in section III that for some canonical geometries $\overline{Z}_{int}$ can be calculated in closed-analytical form. Otherwise, it can always be numerically calculated by solving the interior Dirichlet problem (9). It is also interesting to refer that when the host medium contains several non-connected inclusions in the unit cell the total internal impedance is the sum of the individual internal impedances.

Now that the effective permeability of the metamaterial is characterized, let us examine the effective permittivity problem (with electric field in the *xoy* plane). To this end, we evaluate the electric dipole moment $\mathbf{p}_e$ of the inclusion in the unit cell (per unit of length). From the definition, we have that,

$$\mathbf{p}_e \equiv \int_\Omega (\varepsilon - \varepsilon_h) \mathbf{E}\, ds = \frac{1}{j\omega} \left( \int_\Omega \nabla H_z\, ds \right) \times \hat{\mathbf{u}}_z = 0 \tag{13}$$

where the first identity is a consequence of $\varepsilon_h = 0$, while the second identity follows from the periodicity of the magnetic field. Equation (13) implies that the effective permittivity of the periodic medium is zero at the plasma frequency of the host medium. This result is somehow surprising because one could intuitively expect that by loading a ENZ host with high permittivity inclusions the effective permittivity could be made positive. However, at least in the $\varepsilon_h = 0$ lossless limit, that is not case. A simple justification for this property is that at $\omega = \omega_p$ the fields cannot support phase variations since the magnetic field is necessarily constant in the host medium. Consequently, the effective permittivity can never be positive at $\omega = \omega_p$, because otherwise the metamaterial would support a propagating electromagnetic mode with non vanishing phase variation, which as discussed above is forbidden.

In summary, in this section we proved that at $\omega = \omega_p$ the metamaterial under study is characterized by an effective permittivity $\varepsilon_{eff} = 0$ and an effective permeability given by (12). In the lossless case, the effective permeability is always a real number that can be either positive, negative or zero. An interesting consequence of our theory is that when $\mu_{eff} > 0$ at $\omega = \omega_p$, there is a band gap for frequencies slightly below $\omega_p$, and there is right-handed propagation for frequencies slightly above $\omega_p$ (note that if absorption is absent $\varepsilon_{eff}$ and $\mu_{eff}$ must increase with frequency [9]). Conversely, if $\mu_{eff} < 0$ at $\omega = \omega_p$, there is a band gap for frequencies slightly above $\omega_p$, and there is left-handed propagation for frequencies slightly below $\omega_p$. Finally, if $\mu_{eff} = 0$ at $\omega_p$ the medium is matched and in that case there is left/right-handed propagation for frequencies slightly below/above $\omega_p$, respectively.

## III. Application and Discussion of the Results

To shed some light on the results of the previous section and to gain some intuitive insights, let us consider the particular case in which the cross-section of the basic inclusion is circular with radius $R$, and its permittivity is uniform (i.e. $\varepsilon_i = const.$). In this case, the solution of (9) is $H_z = H_z^{ext} J_0(k_i r)/J_0(k_i R)$, where $k_i = \omega\sqrt{\varepsilon_i \mu_0}$ and $J_l$ is the Bessel function of $1^{st}$ kind and order $l$. The electric field inside the cylinder is:

$$\mathbf{E}^{int} = H_{z,ext} \frac{k_i}{-i\omega\varepsilon_i} \frac{J_1(k_i r)}{J_0(k_i R)} \hat{\mathbf{u}}_\varphi \qquad (14)$$

Note that the electric field lines are azimuthal, and thus a strong magnetic dipole moment may be induced. Using (10) it is straightforward to find the internal impedance of the rod:

$$\bar{Z}_{int} = \frac{2\pi}{A_{cell}} \frac{k_i R}{i\omega\varepsilon_i} \frac{J_1(k_i R)}{J_0(k_i R)} \qquad (15)$$

Substituting this formula into (12) we obtain the effective permeability at the plasma frequency of the host medium,

$$\mu_{eff} = \mu_0 \left( \frac{A_{h,cell}}{A_{cell}} + \frac{2\pi R^2}{A_{cell}} \frac{1}{k_i R} \frac{J_1(k_i R)}{J_0(k_i R)} \right), \qquad \text{at } \omega = \omega_p \qquad (16)$$

where $A_{h,cell} = A_{cell} - \pi R^2$. The above formula is exact: the only assumption is that the permittivity of host material vanishes at $\omega = \omega_p$.

In this work, we are particularly interested in the design of matched zero-index metamaterials with both $\varepsilon_{eff} = 0$ and $\mu_{eff} = 0$. The roots ($k_i R$) of the equation $\mu_{eff} = 0$ (with $\mu_{eff}$ given by (16)) occur near the zeros of the $J_0$-Bessel function. So the first root is around $k_i R \sim 2.405$. To illustrate the possibilities let us consider that the dielectric rods are arranged in a square lattice, with lattice constant $a$. In Fig. 2 we plot the value of $k_i a = \dfrac{\omega_p a}{c} n_i$ (which for a given $\omega_p$ only depends on the index of refraction, $n_i$, of the cylinders) necessary to guarantee that $\mu_{eff} = 0$ as a function of $R/a$. It is seen that the required index of refraction $n_i$ increases very fast when $R/a$ approaches zero. The most favorable situation to design a matched zero-index material is when $R/a \sim 0.5$.

So far the discussion has been restricted to the case in which $\varepsilon_h(\omega_p) = 0$. For obvious reasons it is important to study the dependence of the effective parameters with frequency. Unfortunately, when $\varepsilon_h \neq 0$ the problem cannot be solved analytically as before. Next, we obtain approximate formulae for the effective parameters using the local field approach. To that end, we need to calculate the electric polarizability, $\alpha_e$, and the magnetic polarizability, $\alpha_{m,zz}$, of the inclusion. For the case of a circular rod, it is well-known that the electric polarizability (per unit of length; electric field in the $x$-$y$ plane) is,

$$\alpha_e = \frac{\varepsilon_i - \varepsilon_h}{i \, k_h \varepsilon_i} \pi R \, J_1(k_i R) b_1 \tag{17}$$

with $b_1$ given by

$$b_1 = \left[ -\frac{\pi}{4} \left( J_1(k_i R) H_1^{(1)'}(k_h R) k_h R - \frac{\varepsilon_h}{\varepsilon_i} k_i R \, J_1'(k_i R) H_1^{(1)}(k_h R) \right) \right]^{-1} \tag{18}$$

In above we put $k_h = \omega\sqrt{\varepsilon_h \mu_0}$ and $H_l^{(1)} = J_l + i Y_l$ is the Hankel function of order $l$. In case of dilute systems we have the approximate result:

$$\alpha_e^{-1} \approx \alpha_{e0}^{-1} - i\frac{k_h^2}{8} \qquad \text{where} \quad \alpha_{e0} = \frac{\varepsilon_i - \varepsilon_h}{\varepsilon_i + \varepsilon_h} 2\pi R^2 \qquad (19)$$

The Clausius-Mossotti formula [10] for a square lattice yields that:

$$\varepsilon_{eff} \approx \varepsilon_h \left(1 + \frac{1}{A_{cell}} \frac{1}{\text{Re}\left\{\alpha_e^{-1}\right\} - C_{\text{int},\parallel}}\right) \quad \text{where} \quad C_{\text{int},\parallel} \approx \frac{1}{2}\frac{1}{A_{cell}} \qquad (20)$$

On the other hand, the magnetic polarizability (per unit of length) is,

$$\alpha_{m,zz} = \frac{\varepsilon_i - \varepsilon_h}{\varepsilon_i} \pi R^2 J_2\left(k_i R\right) b_0 \qquad (21)$$

$$b_0 = \left[-i\frac{\pi}{2}\left(-J_0\left(k_i R\right) H_1^{(1)}\left(k_h R\right) k_h R + \frac{\varepsilon_h}{\varepsilon_i} k_i R\, J_1\left(k_i R\right) H_0^{(1)}\left(k_h R\right)\right)\right]^{-1} \qquad (22)$$

and the effective permeability is:

$$\mu_{eff} \approx \mu_0 \left(1 + \frac{1}{A_{cell}} \frac{1}{\text{Re}\left\{\alpha_{m,zz}^{-1}\right\} - C_{\text{int},zz}}\right) \approx \mu_0 \left(1 + \frac{1}{A_{cell}} \frac{1}{\text{Re}\left\{\alpha_{m,zz}^{-1}\right\}}\right) \qquad (23)$$

The second identity follows from the fact that the $z$-directed component of the interaction constant, $C_{\text{int},zz}$, vanishes in the static limit. To illustrate the application of the proposed homogenization formulas, let us suppose that $R = 0.4a$, and that the host medium follows a lossless Drude type model with $\varepsilon_h = \varepsilon_0\left(1 - \omega_p^2/\omega^2\right)$ and $\dfrac{\omega_p}{c}a = \dfrac{2\pi}{3}$. Solving the equation $\mu_{eff} = 0$ we obtain the solution $k_i a = 7.38$ at $\omega = \omega_p$, and consequently the required permittivity for the rods is $\varepsilon_i = 12.4\varepsilon_0$. In Fig. 3 we plot the effective permittivity (full line) and the effective permeability (dashed line) as a function of frequency for this set of parameters. It is seen that the local field theory is consistent with the theory derived previously, and that at $\omega = \omega_p$ both the effective permittivity and permeability vanish. Moreover, it is seen that for $\omega$ slightly smaller than $\omega_p$ the medium is left-handed, while for $\omega$ slightly larger than $\omega_p$ the medium is right-handed, consistently with the observation made in the final of section II. This example shows that in order to ensure that both $\varepsilon_{eff}$ and $\mu_{eff}$ vanish, the required permittivity for the dielectric rods may be large, especially if the electrical size (in free-space units) of the unit cell is small. This is an inconvenience, and so it is interesting to investigate other geometries for the basic inclusion.

Let us consider the two-shell ring depicted in the inset of Fig. 4. The inner shell is defined by $0 < r < R_1$ and is filled with the same plasmonic material as the host. The outer shell $R_1 < r < R_2$ is non-uniform and has permittivity $\varepsilon_i = \varepsilon_i(\varphi)$ ( $r$ and $\varphi$ form a system of polar coordinates relative to the center of the particle; the geometry of Fig. 4 corresponds to the case in which the permittivity $\varepsilon_i = \varepsilon_i(\varphi)$ only assumes two different values). Within the approximation that the particle is electrically small, the second term in the left-hand side of (9) can be neglected and the solution of the equation at $\omega = \omega_p$ subject to boundary condition $H_z = H_z^{ext}$ at $r = R_2$ is:

$$H_z \approx \frac{H_z^{ext} \ln\left(\dfrac{r}{R_1}\right) - H_z^{int} \ln\left(\dfrac{r}{R_2}\right)}{\ln\left(\dfrac{R_2}{R_1}\right)}, \qquad R_1 < r < R_2 \qquad (24)$$

where $H_z^{\text{int}}$ is the (unknown) magnetic field for $r < R_1$ (which is constant because the permittivity of this region vanishes at the plasma frequency). The corresponding electric field is:

$$\mathbf{E}^{\text{int}} \approx \frac{H_z^{\text{int}} - H_z^{\text{ext}}}{\ln\left(\dfrac{R_2}{R_1}\right)} \frac{1}{-i\omega\varepsilon_i} \frac{1}{r} \hat{\mathbf{u}}_{\varphi}, \qquad R_1 < r < R_2 \qquad (25)$$

To calculate the unknown $H_z^{\text{int}}$ we apply Faraday's law to the contour $r = R_1$:

$\displaystyle\oint_{r=R_1} \mathbf{E}^{\text{int}}.\mathbf{dl} = +i\omega\mu_0 H_z^{\text{int}} A_{h,in}$, where $A_{h,in} = \pi R_1^2$. Substituting these results into (10) it is found after straightforward calculations that the internal impedance of the ring can be written as:

$$\bar{Z}_{\text{int}} \approx \frac{1}{\dfrac{1}{-i\omega L_{in}} - i\omega C} \qquad (26)$$

$$L_{in} = \mu_0 \frac{A_{h,in}}{A_{cell}} \qquad ; \qquad C = \varepsilon_{\parallel} \frac{1}{2\pi} \ln\left(\frac{R_2}{R_1}\right) A_{cell} \qquad (27)$$

$$\frac{1}{\varepsilon_{\parallel}} = \frac{1}{2\pi} \int_0^{2\pi} \frac{1}{\varepsilon_i(\varphi)} d\varphi \qquad (28)$$

Thus, $\bar{Z}_{\text{int}}$ is the shunt association of the inductor $L_{in}$ [H/m] and the capacitor $C$ [F.m]. The effective permeability of the periodic medium is given by (12):

$$\mu_{eff} \approx \frac{1}{-i\omega}\left(-i\omega L_{ext} + \frac{1}{\dfrac{1}{-i\omega L_{in}} - i\omega C}\right), \qquad (29)$$

where $L_{ext} = \mu_0 \dfrac{A_{h,ext}}{A_{cell}}$ and $A_{h,ext} = A_{cell} - \pi R_2^2$. In order that the effective permeability vanishes, it is necessary that,

$$k_0^2 \frac{1}{\dfrac{1}{A_{h,in}} + \dfrac{1}{A_{h,ext}}} = \frac{2\pi}{\ln\left(\dfrac{R_2}{R_1}\right)} \frac{\varepsilon_0}{\varepsilon_\parallel} \tag{30}$$

where $k_0 = \dfrac{\omega}{c}$. The above condition can always be fulfilled, no matter how small the particle is, if the profile $\varepsilon_i = \varepsilon_i(\varphi)$ can be chosen in such a way that $\varepsilon_0 / \varepsilon_\parallel \approx 0^+$. If the particle is made of a regular dielectric material this requires that $\varepsilon_i >> \varepsilon_0$, which is not very interesting. However, if low-loss materials with negative permittivity are available, the condition can be satisfied with $|\varepsilon_i|$ having the same magnitude as $\varepsilon_0$, but still resulting in $\varepsilon_0 / \varepsilon_\parallel \approx 0^+$. This situation is potentially interesting at infrared and optical frequencies. Moreover, at microwaves, metals can exhibit high negative permittivity, and therefore the use of a "swiss roll"/split-ring configuration [11] can be considered.

To examine these possibilities, next we consider two examples. We assume that the geometry of the basic inclusion is as depicted in the inset of Fig. 4. The ring is formed by two materials with permittivities $\varepsilon_1$ and $\varepsilon_2$ and filling fractions $f_1$ and $f_2$, respectively, so that $\dfrac{1}{\varepsilon_\parallel} = \dfrac{f_1}{\varepsilon_1} + \dfrac{f_2}{\varepsilon_2}$. As referred before the inner region, $0 < r < R_1$, is filled with an ENZ material. Consider the square lattice formed by an infinite set of these rings embedded in the same ENZ host material as in the previous example. The dimensions of the rings are $R_1 = 0.3a$ and $R_2 = 0.4a$, where $a$ is the lattice constant. In the first example, we suppose that $\varepsilon_1 = -5.0\varepsilon_0$ and that $f_1 = f_2 = 0.5$. In Fig. 4 we plot the effective permeability of the periodic medium as a function of $\varepsilon_2$ (dashed line). It is seen that the effective permeability vanishes around $\varepsilon_2 = 3.7\varepsilon_0$, i.e. when the absolute value of the permittivity of the rings is relatively close. Similar results are obtained even if the electrical size of the rings is very small, and thus this topology can be an interesting option to synthesize matched zero index metamaterials when low loss dielectrics with negative permittivity are available. In the second example, we analyze a split-ring configuration that can be potentially useful at microwaves. Now we suppose that $\varepsilon_1 = -\infty$, i.e. part of the ring is filled with a perfect electric conductor (PEC), and that $f_1 = 1 - f_2 = 0.9$. The solid line in Fig. 4 shows the effective permeability of the structure as a function of $\varepsilon_2$ (dashed line). It is seen that the permeability vanishes when the permittivity of the dielectric gap is $\varepsilon_2 = 2.7\varepsilon_0$. Notice that this value is much smaller than the permittivity required for a dielectric rod with the same size. Thus, as could be expected, the magnetic response of the split ring particle is stronger than that of a dielectric rod.

## IV. Metamaterial Block in a Waveguide Scenario

In the last part of this paper, we will investigate the transmission of a wave through a finite sample of the metamaterial under study. We consider the setup shown in Fig. 5, which consists of two parallel-plate waveguide sections. The walls of the waveguide are made of perfect electric conductors (PEC). The transition region between the waveguides is filled with a metamaterial having an ENZ host with dielectric inclusions. Let us consider that the fundamental transverse electromagnetic (TEM) mode impinges on $x$=0 interface, as depicted in Fig. 5. In [6] it was proven that if there are no inclusions in the host medium and for $\varepsilon_h(\omega) = 0$, the reflection coefficient at the input interface is exactly given by (independent of the geometry of the plasmonic transition region):

$$\rho = \frac{d_1 - d_2 + ik_0\mu_{r,p}A_{ch}}{d_1 + d_2 - ik_0\mu_{r,p}A_{ch}} \tag{31}$$

where $A_{ch}$ is the area of the ENZ channel, and $\mu_{r,p}$ is the relative permeability of the host. In the particular case in which the cross-section of the two parallel plate waveguides is the same, $d_1 = d_2$, the reflection coefficient can be made arbitrarily small if the permeability $\mu_{r,p}$ of the ENZ material can be adjusted so that it approaches zero. This can be useful to improve the efficiency of waveguiding devices at junctions and bends. Based on this idea, we will try to design a metamaterial made of a non-magnetic ENZ host loaded with dielectric inclusions. The inclusions are designed in such a way that the effective permeability is near zero. Let us then consider that the ENZ host is loaded with $N_i$ inclusions, as illustrated in Fig. 5. For convenience we suppose that all the inclusions are identical, and that the cross-section of a generic inclusion is $D$, even though that assumption is not essential. The theory developed in [6] can be easily generalized to this geometry. Detailed calculations show that at $\omega = \omega_p$ the reflection coefficient is given by,

$$\rho = \frac{d_1 - d_2 + \left( ik_0 A_h + \dfrac{N_i}{\eta_0 H_z^{ext}} \oint_{\partial D} \mathbf{E}^{\text{int}} \cdot \mathbf{dl} \right)}{d_1 + d_2 - \left( ik_0 A_h + \dfrac{N_i}{\eta_0 H_z^{ext}} \oint_{\partial D} \mathbf{E}^{\text{int}} \cdot \mathbf{dl} \right)} \tag{32}$$

where $A_h$ is the area of the host material (area of the channel, $A_{ch}$, subtracted from the area of the inclusions), $\eta_0 = \sqrt{\mu_0/\varepsilon_0}$ is the free-space impedance, and $\mathbf{E}^{\text{int}}$ is defined as in section II, i.e. it is the electric field associated with the solution of the boundary value problem defined by equation (9) with $H_z = H_z^{ext}$ at $\partial D$. If we regard the waveguide transition as a metamaterial as in the first part of this paper, it is obvious that the area of the corresponding unit cell is $A_{cell} = A_{ch}/N_i$. Thus, using (10) we can rewrite (32) as:

$$\rho = \frac{d_1 - d_2 + \left( ik_0 A_h - \dfrac{1}{\eta_0} A_{ch} \bar{Z}_{\text{int}} \right)}{d_1 + d_2 - \left( ik_0 A_h + \dfrac{1}{\eta_0} A_{ch} \bar{Z}_{\text{int}} \right)} \tag{33}$$

Using (12) and identifying $A_{h,cell} = A_h/N_i$, the formula simplifies to:

$$\rho = \frac{d_1 - d_2 + ik_0 A_{ch} \dfrac{\mu_{eff}}{\mu_0}}{d_1 + d_2 - ik_0 A_{ch} \dfrac{\mu_{eff}}{\mu_0}} \tag{34}$$

where $\mu_{eff}$ is defined consistently with the theory of section II. The above result is exact, at $\omega = \omega_p$. This is remarkable because the geometry of the plasmonic channel and the inclusions can be completely arbitrary. Moreover, comparing (34) with (31) the formula is even more surprising: it is seen that a ENZ channel filled with the discrete non-magnetic dielectric inclusions is equivalent to a channel filled with an ideal continuous ENZ material with $\mu_{r,p} = \mu_{eff}/\mu_0$. That is, at the plasma frequency and in the described waveguide scenario, the incoming wave cannot distinguish if the channel is filled with a continuous medium or with a metamaterial implementation of the medium. Note that (34) is valid for an arbitrary number of inclusions ($N_i \geq 1$) and the inclusions may be arbitrarily positioned, i.e. they do not have to be arranged in a regular lattice. This means that the electromagnetic fields do not "see" the granularity of the metamaterial (i.e. its non-uniform nature) at the plasma frequency, $\omega = \omega_p$. Possibly the reason is that since the wave number in the host medium is zero, the inclusions look always electrically small, independently of their actual physical size. An important corollary of these properties is that in opposition to the traditional theory, a metamaterial with an ENZ host can be homogenized (at least over some frequency band) even if the electrical size of the inclusions is large. Indeed, the ENZ host forces the inclusions to behave as lumped elements characterized by a certain internal impedance $\overline{Z}_{\text{int}}$, independently of its actual electrical size in free-space unities.

To study the suggested possibilities and the effect of losses in a realistic setup, let us consider the geometry depicted in Fig. 6, with a channel formed by two 90-[deg] bends filled with a metamaterial. The distance between the parallel plates is $d$. In the first example, we assume that the inclusions have circular cross-section with radius $R = 0.4a$. It was seen in section III that in order to have $\mu_{eff} = 0$ it is necessary that $k_i a = 7.38$ at $\omega = \omega_p$. Assuming that the inclusions are arranged in a square lattice, the lattice constant is given by $a = d/\sqrt{N_i/4}$ (we assume that $N_i$ is multiple of 4 because the area of the channel is $A_{ch} = 4d^2$). Thus the permittivity required for the rods is $\varepsilon_i = 12.4\frac{N_i}{4}\varepsilon_0$, i.e. the permittivity increases linearly with the number of inclusions. The reason is that a single rod can induce a magnetic dipole moment larger than that of many small rods with comparable total area. Thus, the most interesting case is the one in which the unit cell has a small number of inclusions. We will assume that $N_i = 4$, which corresponds to the geometry depicted in Fig. 6. To study the effect of losses and frequency dispersion we consider that the rods are embedded in an ENZ host that follows the Drude dispersion model $\varepsilon_h = \varepsilon_0 \left( 1 - \dfrac{\omega_p^2}{\omega(\omega + i\Gamma)} \right)$, where the plasma frequency is such that $\dfrac{\omega_p}{c} d = \dfrac{2}{3}\pi$, and

$\Gamma$ is the collision frequency [rad/s]. Note that at $\omega = \omega_p$, we have $\mathrm{Re}(\varepsilon) \approx 0$ and

$\varepsilon / \varepsilon_0 \approx +i \, \Gamma / \omega_p$. Using a finite-integration technique commercial simulator CST

Microwave Studio[TM] [12], we computed the transmission characteristic (s21 parameter)

through the channel for different configurations. The results are depicted in Fig. 7. Curve

a) shows the s21 parameter when the channel is empty, and demonstrates that near

$\omega = \omega_p$ the transmission is very low. When the channel is filled with the ENZ material –

curve b) – the transmission around $\omega = \omega_p$ slightly improves but is still residual. Quite

differently, when the 4 dielectric rods are inserted in ENZ channel, the transmission

around $\omega = \omega_p$ is greatly improved – curves c) and d) – even in case of significant losses

$\Gamma / \omega_p = 0.05$. Note that the channel width is as large as $1.33\lambda_0$ free-space wavelengths at

the design frequency, and thus the effect of losses can be considered somehow moderate.

When losses are negligible the wave completely tunnels through the bend, as illustrated in curve c).

In the second example, we consider again the geometry shown in Fig. 6, except that now the dielectric rods are replaced by split rings with geometry similar to that shown in the inset of Fig. 4. In order to get $\mu_{eff} = 0$ we choose $\varepsilon_1 = -\infty$ (PEC material),

$\varepsilon_2 = 2.7\varepsilon_0$, and $f_1 = 1 - f_2 = 0.1$, consistently with the results obtained in section III. The

host material is the same as in the previous example. The computed transmission characteristic is shown in Fig. 8. Note that curves a) – empty channel - and b) – channel filled with ENZ host - are the same as in the previous example. The set of 3 curves labeled with c) show the transmission characteristic when the ENZ channel is loaded with 4 split ring resonators, assuming $\Gamma / \omega_p = 0$, 0.01, and 0.05, respectively. Again it is seen that in case of small losses the channel is completely matched to the input and output waveguides. Finally, curve d) depicts the transmission characteristic when the channel is loaded with split rings but the ENZ host is replaced by air. As seen no significant transmission is possible in those circumstances around $\omega = \omega_p$.

To conclude this section, we discuss how the theoretical concepts introduced in this paper can be demonstrated in a practical setup at microwaves. At this frequency range ENZ materials are not readily available in nature. However, it has long been known, (see e.g., [13]), that parallel metallic plates (normal to the $z$-direction) can simulate a two-dimensional artificial plasma when the electric field is parallel to the plates. The effective permittivity of the parallel-plate medium follows the Drude-type model $\varepsilon_h / \varepsilon_0 = \varepsilon_{d,r} - \left( \pi / k_0 s \right)^2$, where $\varepsilon_{d,r}$ is the relative permittivity of the dielectric between the plates, and $s$ is the distance between the plates. The region $0 < z < s$ delimited by the metallic plates behaves effectively as a 2D-plasmonic medium for propagation along the $xoy$ plane. We note that this concept was used in [14] to demonstrate that a set of split-ring resonators in a waveguide environment support subwavelength left-handed propagation.

Based on this artificial plasma concept, how can we design a 3D-configuration that emulates the 2D-waveguide scenario depicted in Fig. 6? In order that the setup can be tested experimentally it is interesting to study a configuration that corresponds to a closed 3D-waveguide environment. But in the configuration of Fig. 6 some regions are filled with air. How can we overcome this obstacle?

First let us discuss how to design the ENZ artificial material. We suppose that in the channel region the plates are filled with air. Thus, in order that $\frac{\omega_p}{c}d = \frac{2}{3}\pi$ it is necessary that the distance between the plates is $s = 1.5d$. The next important issue is how to emulate the free-space regions in the configuration of Fig. 6 (Region 1 and Region 2). As referred above, the permittivity of these regions must be that of free-space. But this can be easily achieved at $\omega = \omega_p$, if the plates are filled with a dielectric with permittivity $2.0\varepsilon_0$ in those regions. That is, around $\omega = \omega_p$ the artificial plasma (formed by two parallel metallic plates) behaves as an ENZ material in the channel region, and as free-space in regions 1 and 2 because in these regions the dielectric spacer has higher permittivity. We can design the dielectric inclusions using similar arguments. More specifically, to emulate a material with permittivity $\varepsilon_i$ in the environment shown in Fig. 6 , we need to load the artificial plasma with a material with corrected permittivity $\varepsilon_i + \varepsilon_0$. Using these ideas we obtain the configuration shown in Fig. 9, which simulates the behavior of the 2D-structure shown in Fig. 6 around $\omega = \omega_p$, except that we consider that the inclusions are split ring cylinders rather than dielectric rods. All the walls are formed by PEC materials, and so the proposed geometry corresponds to a closed metallic waveguide environment. The split rings are designed using the same parameters as in the second example of this section. Thus, the permittivity of the dielectric gap is now $\varepsilon_2 = (2.7 + 1)\varepsilon_0$. Using CST Microwave Studio[TM] [12], we computed the transmission characteristic of the proposed configuration in different scenarios. The waveguide is excited with the fundamental TE10 mode with electric field in the xoy-plane. This mode emulates the behavior of the TEM mode in configuration of Fig. 6 around $\omega \approx \omega_p$. The TE10 mode can propagate for frequencies larger than $\omega > 0.7\omega_p$ (note that around $\omega = 0.7\omega_p$ regions 1 and 2 behave as ENZ, while region 3 simulates a material with negative permittivity). In Fig. 10 we plot the S21 parameter for the cases *a*) the ring inclusions are removed and all regions are loaded with a material with permittivity $2.0\varepsilon_0$, *b*) the ring inclusions are removed, region 1 and 2 are loaded with a dielectric with permittivity $2.0\varepsilon_0$ and region 3 is filled with air, *c*) the ring inclusions are considered, region 1 and 2 are loaded with a dielectric with permittivity $2.0\varepsilon_0$ and region 3 is filled with air. Note that configuration a) is expected to emulate the behavior of an empty channel in Fig. 6, configuration b) is expected to emulate the behavior of ENZ unloaded channel, while configuration c) is expected to simulate an ENZ channel loaded with rings. In fact, comparing curves a), b) and c) in Fig. 10 with the corresponding curves in Fig. 8, one sees that all curves have a very similar frequency dependence around $\omega \approx \omega_p$. This remarkable result demonstrates that metallic waveguides can simulate the matched zero-index metamaterials under study. Notice that in configuration *c*) the wave completely tunnels through the plasmonic channel, as predicted by our theory. To conclude we note that when the frequency is far from $\omega_p$ the agreement between Fig. 8 and Fig. 10 is not so good, because in the artificial plasma configuration the effective permittivity of regions 1 and 2 is not anymore close to that of free-space.

## V. Conclusions

In this work, the electrodynamics of metamaterials with an ENZ host was studied. A general and rigorous homogenization approach to characterize these artificial materials was derived. We obtained a closed-form expression for the effective permeability of the metamaterials, and proved that for some canonical inclusion shapes the permeability can be calculated in closed analytical form. The possibility of designing matched zero-index materials was studied and demonstrated numerically. In particular, we investigated the response of a finite sized metamaterial block in a complex waveguide scenario. It was shown that an incoming wave cannot distinguish between a continuous medium and a metamaterial with ENZ host, i.e., the wave cannot see the granularity of the metamaterial. Also it was proved that unlike other artificial media, the structures investigated here can be homogenized over some frequency band even if the size of the inclusions is large when compared with the wavelength. To study the effect of losses and of frequency dispersion we calculated the transmission of an incoming wave through a bend filled with a properly designed metamaterial. We proved that the proposed metamaterial structures can be realized in the microwave domain using the concept of artificial plasma.

## Acknowledgments:

This work is supported in part by the U.S. Air Force Office of Scientific Research (AFOSR) grant number FA9550-05-1-0442. Mário Silveirinha has been partially supported by a fellowship from "Fundação para a Ciência e a Tecnologia".

# Appendix A

Here, we discuss the properties of the electromagnetic Floquet-Bloch modes supported by the metamaterial studied in section II. We assume that at $\omega = \omega_p$ the permittivity of the host vanishes: $\varepsilon_h \left( \omega_p \right) = 0$.

To begin with, we note that the fundamental result (1) implies that an electromagnetic mode associated with the wave vector $\mathbf{k} = (k_x, k_y, 0)$, i.e. such that $H_z(\mathbf{r})\exp(j\mathbf{k}.\mathbf{r})$ is periodic, can only exist at the frequency $\omega = \omega_p$ if either $\mathbf{k} = \mathbf{0}$ or if $H_z^{ext} = 0$. It can be verified that the latter condition implies that $H_z = 0$ in the whole crystal and that the electric field vanishes inside the inclusions. However, the electric field in the host medium is different from zero. In fact, it can be proven that for each $\mathbf{k} \neq 0$ there exists an electromagnetic mode with these properties, and, moreover, that the polarization of the mode is longitudinal i.e. the average electric field is directed along $\mathbf{k}$. This set of longitudinal modes is the perfect analogue of the set of longitudinal modes characteristic of a cold plasma at the plasma frequency. Notice that the magnetic field associated with the longitudinal modes is exactly zero. In addition to this set of longitudinal modes, the periodic medium also supports electromagnetic modes associated with $\mathbf{k} = \mathbf{0}$. These solutions form an infinite family of periodic modes with $H_z = 0$. Indeed, for every vector $\mathbf{E}_{av}$ it is possible to find a periodic field $\mathbf{E}_{p0}$ such that it satisfies $\nabla \times \mathbf{E}_{p0} = 0$ in the host medium, its tangential component vanishes at the boundary of the inclusion $\partial D$, and the spatial average of $\mathbf{E}_{p0}$ equals $\mathbf{E}_{av}$. Note that $\mathbf{E}_{p0}$ is not necessarily unique because in the $\varepsilon_h = 0$ limit the divergence of $\mathbf{E}_{p0}$ is not necessarily zero in the ENZ medium (nevertheless, a divergence free solution always exists).

In addition to the set of modes described above, the periodic medium also supports a set of *generalized* Floquet modes with electric field of the form $\mathbf{E} = \dfrac{x_i}{a}\mathbf{E}_{p0} + \mathbf{E}_{p1}$, as detailed in section II. The mathematical argument that justifies the emergence of these modes is also sketched in section II.

*Figures*

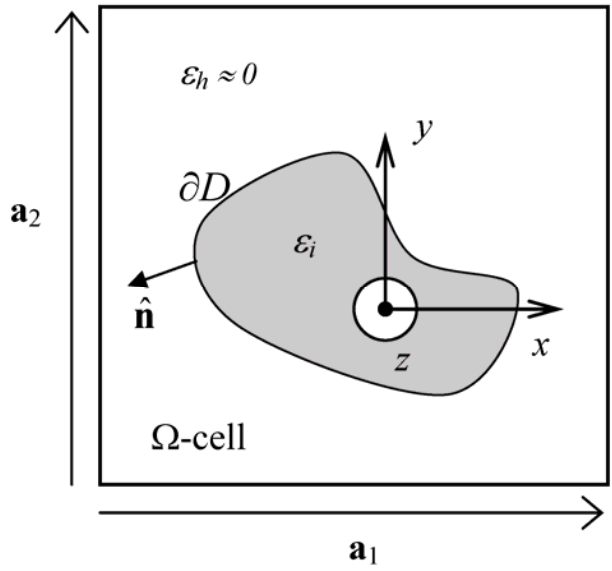

Fig. 1. Geometry of the unit cell-Ω of the metamaterial crystal (generic configuration). The permittivity of the host medium is $\varepsilon_h$ and the permittivity of the non-magnetic inclusion is $\varepsilon_i$. The permeability of each material is $\mu_o$.

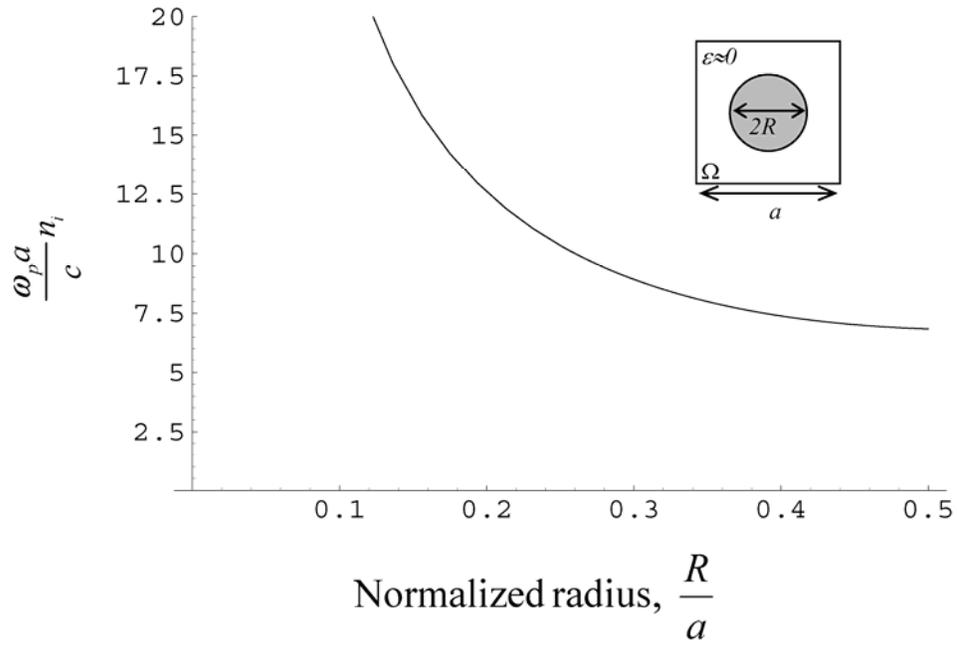

Fig. 2 Plot of the (normalized) index of refraction of the rods that guarantees $\mu_{eff} = 0$, as a function of the normalized radius $R/a$. The inset shows the geometry of the unit cell.

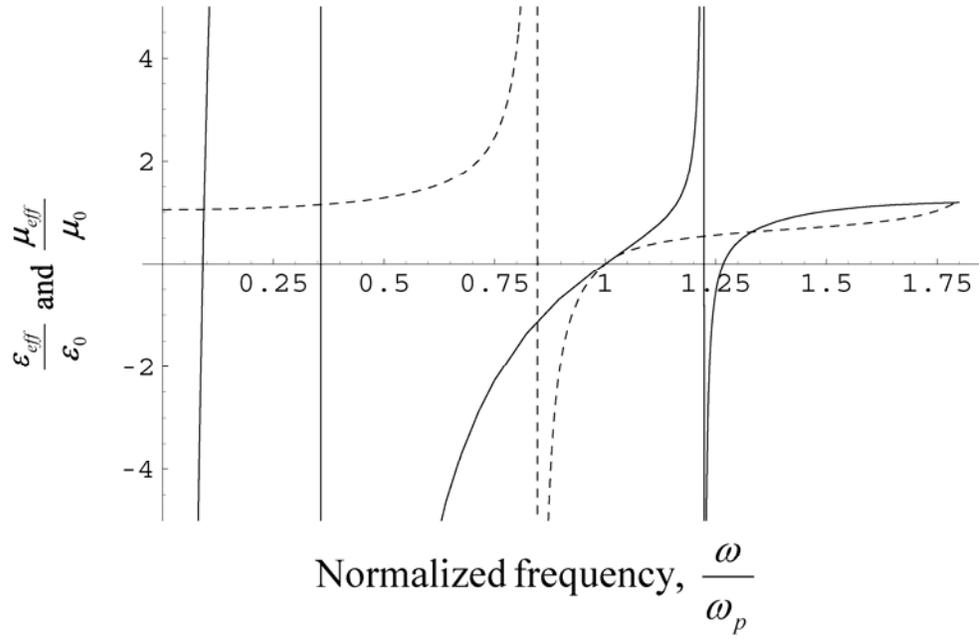

Fig. 3 Plot of $\varepsilon_{eff}/\varepsilon_0$ (solid line) and $\mu_{eff}/\mu_0$ (dashed line) as a function of $\omega/\omega_p$. The metamaterial

consists of a square array of cylindrical rods with permittivity $\varepsilon_i = 12.4\varepsilon_0$ and radius $R = 0.4a$. The rods are embedded in a host medium with a Drude permittivity model such that $\omega_p a/c = 2\pi/3$.

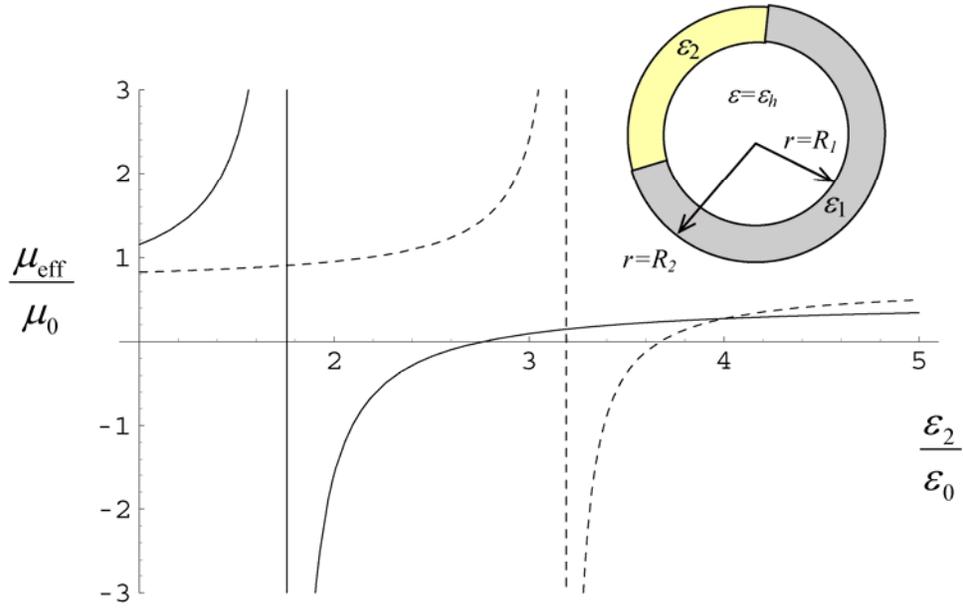

Fig. 4 (Color online) Effective permeability of a metamaterial formed by ring inclusions as a function of $\varepsilon_2$ (assuming $\varepsilon_h \approx 0$). The geometry of the basic inclusion is shown in the inset. The core of the particle, $0 < r < R_1$, is filled with the same material as the host medium. Solid line: $\varepsilon_1 = -\infty$ (PEC material) and $f_1 = 0.9$. Dashed line: $\varepsilon_1/\varepsilon_0 = -5.0$ and $f_1 = 0.5$.

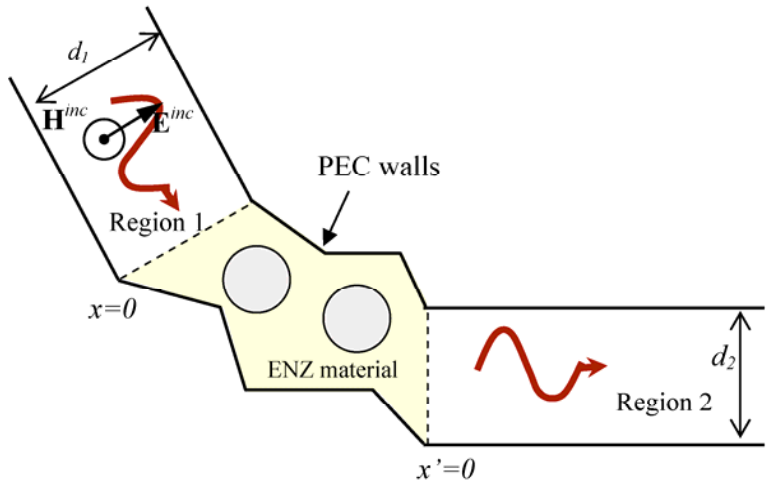

Fig. 5 (Color online) Geometry of a generic two-dimensional (2-D) waveguide structure with an ENZ material section containing some non-magnetic inclusions with circular cross-section. The waveguide walls in regions 1 and 2 are parallel to the $x$ and $x'$-directions, respectively. The interfaces of the ENZ material channel are planar and normal to the waveguide walls.

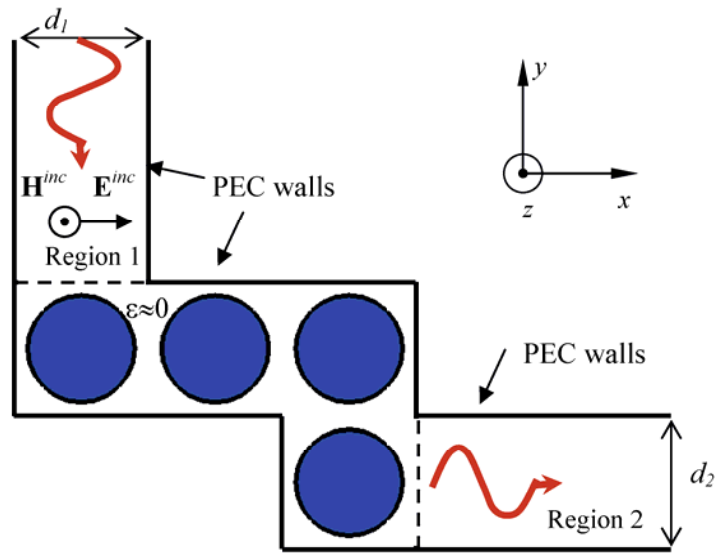

Fig. 6 (Color online) Geometry of parallel-plate waveguides connected through a channel filled with a metamaterial with ENZ host and dielectric inclusions.

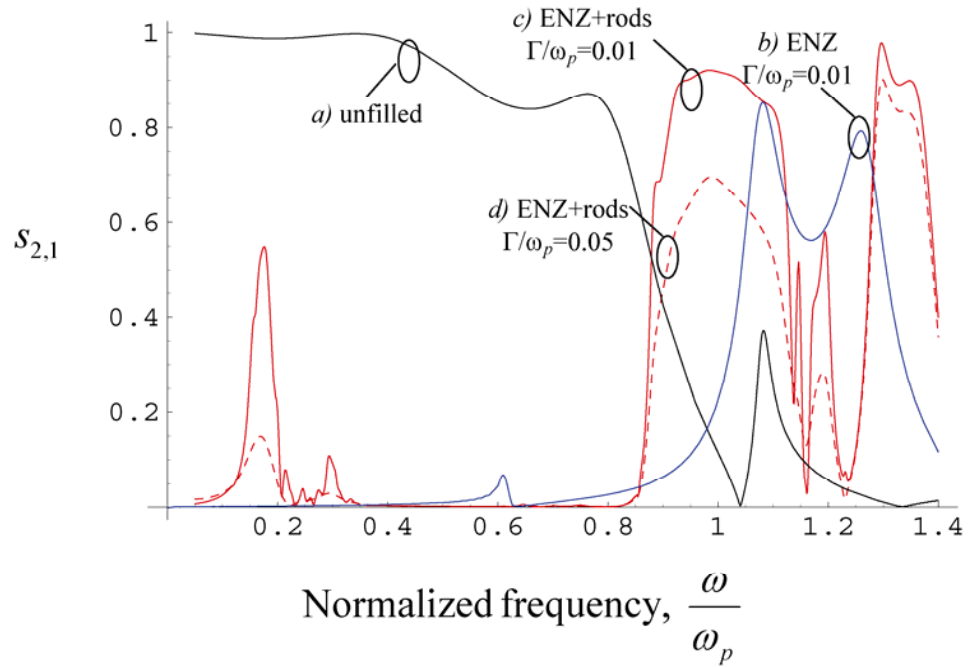

Fig. 7 (Color online) Transmission characteristic (s21) as a function of the normalized plasma frequency. *a)* Unfilled bend. *b)* Bend is filled with ENZ material. *c)* and *d)* bend is filled with an ENZ material ( $\Gamma / \omega_p$ =0.01 and 0.05, respectively) loaded with 4 dielectric rods.

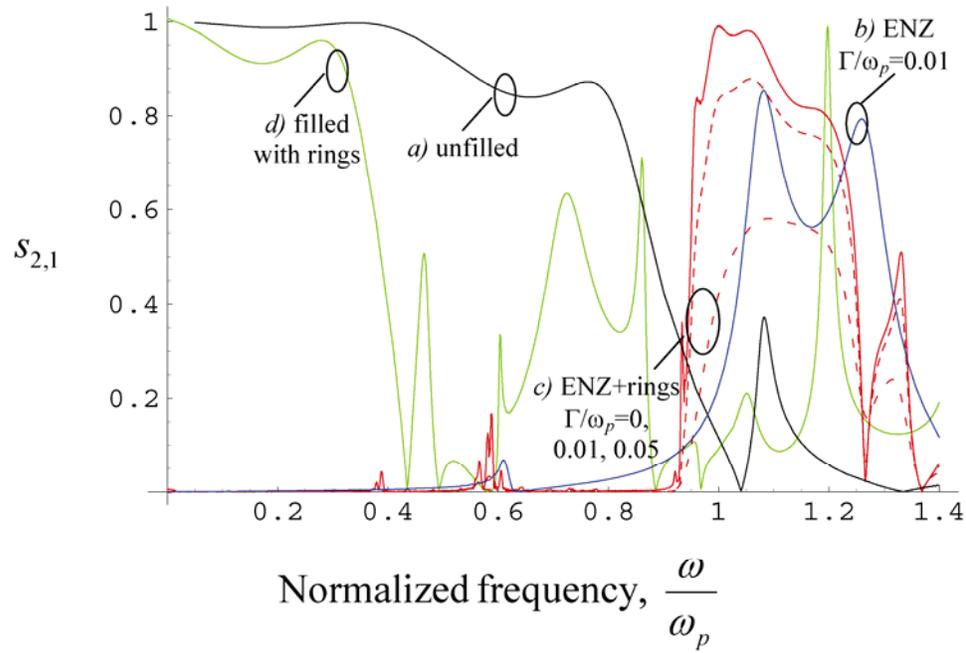

Fig. 8 (Color online) Transmission characteristic (s21) as a function of the normalized plasma frequency. *a)* Unfilled bend. *b)* Bend is filled with ENZ material. *c)* bend is filled with an ENZ material ( $\Gamma/\omega_p =0$, 0.01, 0.05, respectively) loaded with 4 rings. *d)* bend is filled with 4 rings standing in air.

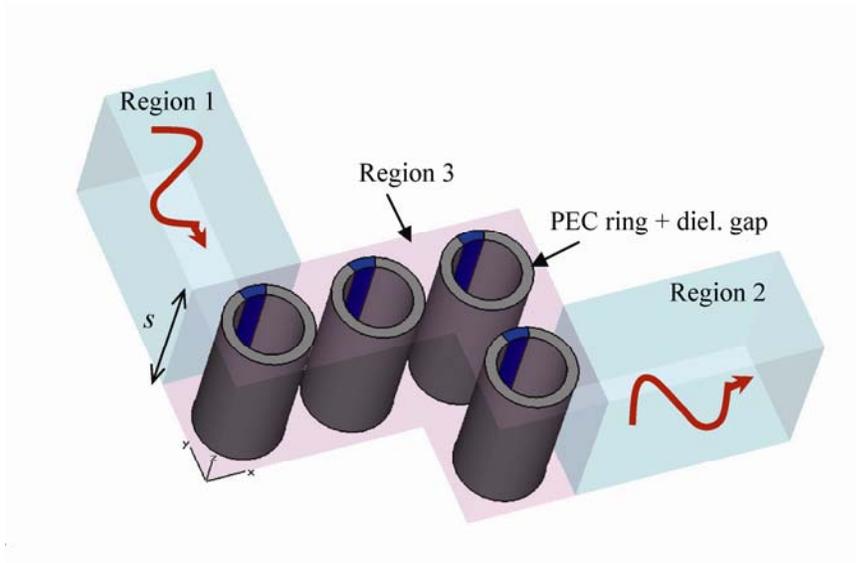

Fig. 9 (Color online) Geometry of the 3D-dimensional rectangular metallic waveguide that emulates the behavior of the 2D-dimensional structure. The H-plane width, *s*, is chosen so that Region 3 behaves as an artificial plasma. Both regions 1 and 2 are filled with a dielectric so that they are operated in a region where their effective index of refraction is unity.

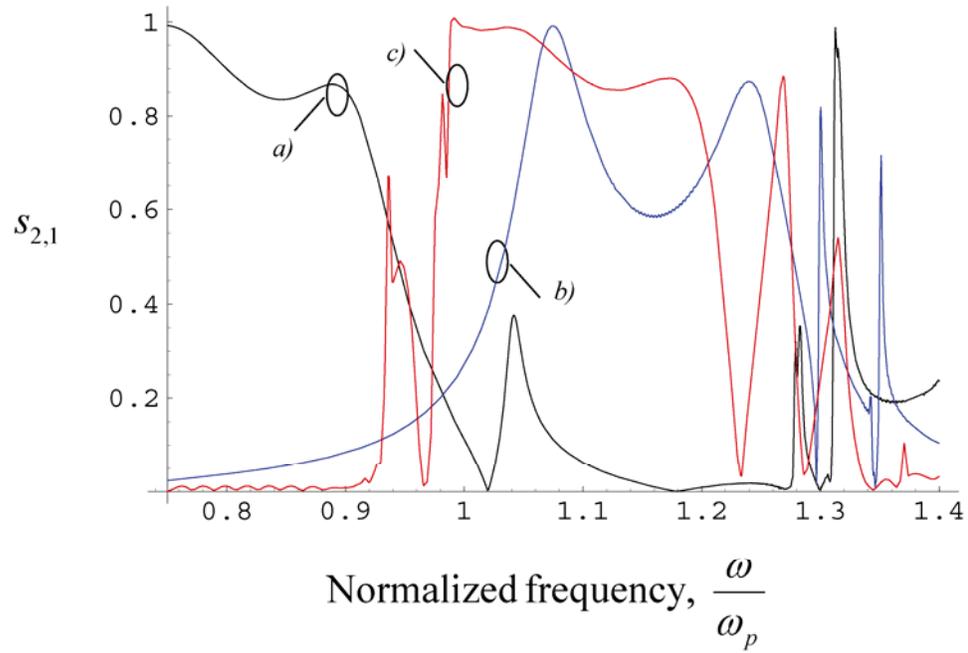

Fig. 10 (Color online) Transmission characteristic (s21) as a function of the normalized plasma frequency for a metamaterial realization that emulates *a)* an unfilled bend. *b)* a bend filled with ENZ material. *c)* a bend filled with ENZ material loaded with ring inclusions.